\documentclass[twocolumn,showpacs,preprintnumbers,amsmath,amssymb]{revtex4}

\usepackage{graphicx}
\usepackage{dcolumn}
\usepackage{bm}

\begin{document}
\draft
\title{Non-Markovian Decay and Lasing Condition in an Optical Microcavity Coupled to a Structured Reservoir}
\normalsize
\author{Stefano Longhi}
\address{Dipartimento di Fisica and Istituto di Fotonica e Nanotecnologie del CNR,
Politecnico di Milano, Piazza L. da Vinci 32,  I-20133 Milan,
Italy}


%
\bigskip
\begin{abstract}
\noindent The decay dynamics of the classical electromagnetic
field in a leaky optical resonator supporting a single mode
coupled to a structured continuum of modes (reservoir) is
theoretically investigated, and the issue of threshold condition
for lasing in presence of an inverted medium is comprehensively
addressed. Specific analytical results are given for a single-mode
microcavity resonantly coupled to a coupled resonator optical
waveguide (CROW), which supports a band of continuous modes acting
as decay channels. For weak coupling, the usual exponential
Weisskopf-Wigner (Markovian) decay of the field in the bare
resonator is found, and the threshold for lasing increases
linearly with the coupling strength. As the coupling between the
microcavity and the structured reservoir increases, the field
decay in the passive cavity shows non exponential features, and
correspondingly the threshold for lasing ceases to increase,
reaching a maximum and then starting to decrease as the coupling
strength is further increased. A singular behavior for the "laser
phase transition", which is a clear signature of strong
non-Markovian dynamics, is found at critical values of the
coupling between the microcavity and the reservoir.
\end{abstract}



\maketitle

\newpage

\section{Introduction.}
It is well known that the modes of an open optical cavity are
always leaky due to energy escape to the outside. Mode leakage can
be generally viewed as due to the coupling of the discrete cavity
modes with a broad spectrum of modes of the "universe" that acts
as a reservoir \cite{Lang73,Ching87,Ching98}. From this
perspective the problem of escape of a classical electromagnetic
field from an open resonator is analogous to the rather general
problem of the decay of a discrete state coupled to a broad
continuum, as originally studied by Fano \cite{Fano64} and
encountered in different physical contexts (see, e.g.,
\cite{Tannoudji}). The simplest and much used way to account for
mode coupling with the outside is to eliminate the reservoir
degrees of freedom by the introduction of quasi normal modes with
complex eigenfrequencies (see, e.g., \cite{Lang73,Ching98}), in
such a way that energy escape to the outside is simply accounted
for by the cavity decay rate $\gamma$ (the imaginary part of the
eigenvalue) or, equivalently, by the cavity quality factor $Q$.
This irreversible exponential decay of the mode into the continuum
corresponds to the well-known Weisskopf-Wigner decay and relies on
the so-called Markovian approximation (see, e.g.,
\cite{Tannoudji}) that assumes an instantaneous reservoir response
(i.e. no memory): coupling with the reservoir is dealt as a
Markovian process and the evolution of the field in the cavity
depends solely on the present state and not on any previous state
of the reservoir. For the whole system (cavity plus outside), in
the Markovian approximation the cavity quasi-mode with a complex
frequency corresponds to a resonance state with a Lorentzian
lineshape. If now the field in the cavity experiences gain due to
coupling with an inverted atomic medium, the condition for lasing
is simply obtained when gain due to lasing atoms cancels cavity
losses, i.e. for $g=\gamma$, where $g$ is the modal gain
coefficient per unit time \cite{Lang73}. More generally, treating
the field classically and assuming that the cavity supports a
single mode, an initial field amplitude in the cavity will
exponentially decay, remain stationary (delta-function lineshape)
or exponentially grow (in the early stage of lasing) depending on
whether $g< \gamma$, $g=\gamma$ or $g> \gamma$, respectively. In
addition, since the cavity decay rate $\gamma$ increases as the
coupling of the cavity with the outside increases, the threshold
for laser oscillation increases as the coupling strength of the
resonator with the modes of the "universe" is increased. It is
remarkable that this simple and widely acknowledged dynamical
behavior of basic laser theory, found in any elementary laser
textbook (see, e.g., \cite{Svelto}), relies on the Markovian
assumption for the cold cavity decay dynamics \cite{note0}.
However, it is known that in many problems dealing with the decay
of a discrete state coupled to a "structured" reservoir, such as
in photoionization in the vicinity of an autoionizing resonance
\cite{Piraux90}, spontaneous emission and laser-driven atom
dynamics in waveguides and photonic crystals
\cite{Lai88,Lewenstein88,John90,John94,Kofman94,Vats98,Lambropoulos00,Wang03,Petrosky05},
and electron transport in semiconductor superlattices
\cite{Tanaka06}, the Markovian approximation may become invalid,
and the precise structure of the reservoir (continuum) should be
properly considered. Non-Markovian effects may become of major
relevance in presence of threshold \cite{Piraux90,Gaveau95} or
singularities
\cite{Lewenstein88,John94,Kofman94,Lambropoulos00,Tanaka06} in the
density of states or more generally when the coupling strength
from the initial discrete state to the continuum becomes as large
as the width of the continuum density of state distribution
\cite{Tannoudji}. Typical features of non-Markovian dynamics found
in the above-mentioned contexts are non-exponential decay,
fractional decay and population trapping, atom-photon bound
states, damped Rabi oscillations, etc. Though the role of
structured reservoirs on basic quantum electrodynamics and quantum
optics phenomena beyond the Markovian approximation has received a
great attention (see, e.g., Ref.\cite{Lambropoulos00} for a rather
recent review), at a classical level  \cite{Ching98} previous
works have mainly considered  the limit of Markovian dynamics
\cite{Lang73}, developing a formalism based on quasi-normal mode
analysis of the open system \cite{Ching98}. In fact, in a typical
laser resonator made e.g. of two-mirrors with one partially
transmitting mirror coupled to the outside open space, the
Weisskopf-Wigner decay law for the bare cavity field is an
excellent approximation \cite{Lang73} and therefore non-Markovian
effects are fully negligible. However, the advent of micro- and
nano-photonic structures, notably photonic crystals (PCs), has
enabled the design and realization of high-$Q$ passive
microcavities
\cite{Villeneuve96,Vahala03,Armani03,Asano04,Asano06} and lasers
\cite{Vahala03,Painter99,Loncar02,Park04,Altug05} which can be
suitably coupled to the outside by means of engineered waveguide
structures \cite{Vahala03,Fan98,Xu00,Asano03,Waks05,Chak06}. By
e.g. modifying some units cells within a PC, one can create
defects that support localized high-$Q$ modes or propagating
waveguide modes. If we couple localized defect modes with
waveguides, many interesting photon transport effects may occur
(see, e.g., \cite{Fan98,Xu00,Fan05}). Coupling between optical
waveguides and high-$Q$ resonators in different geometries has
been investigated in great detail using numerical methods,
coupled-mode equations, and scattering matrix techniques in the
framework of a rather general Fano-Anderson-like Hamiltonian
\cite{Fan98,Xu00,Asano03,Waks05,LanLan05,Chak06}. Another kind of
light coupling and transport that has received an increasing
attention in recent years is based on coupled resonator optical
waveguide (CROW) structures
\cite{Stefanou98,Yariv99,Ozbay00,Olivier01}, in which photons hop
from one evanescent defect mode of a cavity to the neighboring one
due to overlapping between the tightly confined modes at each
defect site. The possibility of artificially control the coupling
of a microcavity with the "universe" may then invalidate the usual
Markovian approximation for the (classical) electromagnetic field
decay. In such a situation, for the passive cavity one should
expect to observe non-Markovian features in the dynamics of the
decaying field, such as non-exponential decay, damped Rabi
oscillations, and quenched decay for strong couplings. More
interesting, for an active (i.e. with gain) microcavity the usual
condition $g=\gamma$ of gain/loss balance for laser oscillation
becomes meaningless owing to the impossibility of precisely define
a cavity decay rate $\gamma$. Therefore the determination of the
lasing condition for a microcavity coupled to a structured
reservoir requires a detailed account of the mode structure of the
universe and may
show unusual features.\\
It is the aim of this work to provide some general insights into
the classical-field decay dynamics and lasing condition of an
optical microcavity coupled to a structured reservoir, in which
 the usual Markovian approximation of treating the cavity decay
 becomes inadequate. Some general results are
 provided for a generic Hamiltonian model describing the
 coupling of a single-mode microcavity with a continuous band of
 modes, and the effects of non-Markovian dynamics on lasing condition are discussed.
 As an illustrative example, the case of a microcavity resonantly coupled to a CROW
is considered, for which analytical results may be given in a closed form.\\
The paper is organized as follows. In Sec.II a simple model
describing the classical field dynamics in an active single-mode
microcavity coupled to a band of continuous modes is presented,
and the Markovian dynamics attained in the weak coupling regime is
briefly reviewed. Section III deals with the exact dynamics,
beyond the Markovian approximation, for both the passive (i.e.
without gain) and active  microcavity. In particular, the general
relation expressing threshold for laser oscillation is derived,
and its dependence on the coupling strength between the
microcavity and the reservoir is discussed. The general results of
Sec.III  are specialized in Sec.IV for the case of a single-mode
microcavity tunneling-coupled to a CROW, and some unusual
dynamical effects (such as "uncertainty" of laser threshold,
non-exponential onset of lasing instability and transient
non-normal amplification) are shown to occur at certain critical
couplings.

\section{Microcavity coupled to a structured reservoir: description of the model and Markovian dynamics}
\subsection{The model}
The starting point of our analysis is provided by a rather general
Hamiltonian model \cite{Fan98,Xu00} describing the interaction of
a localized mode $|a\rangle$ of a resonator system (e.g. a
microcavity in a PC) with a set of continuous modes
$|\omega_{\mu}\rangle$ of neighboring waveguides with which the
resonator is tunneling-coupled. We assume that the microcavity
supports a single and high-$Q$ localized mode of frequency
$\omega_a$,  and indicate by $\gamma_i$ and $g$ the intrinsic
losses and gain coefficients of the mode. The intrinsic losses
$\gamma_i$ account for both internal (e.g. absorption) losses and
damping of the cavity mode due to coupling with a "Markovian"
reservoir (i.e. coupling with modes of the universe other than the
neighboring waveguides). The modal gain parameter $g$ may be
provided by an inverted atomic or semiconductor medium hosted in
the microcavity. Since we will consider the microcavity operating
below or at the onset of threshold for lasing, as in
Refs.\cite{Xu00,LanLan05} the modal gain parameter $g$ is assumed
to be a constant and externally controllable parameter; above
threshold an additional rate equation for $g$ would be obviously
needed depending on the specific gain medium (see, for instance,
\cite{Liu05}). Dissipation and gain of the microcavity mode are
simply included in the model by adding a non-Hermitian term
$H_{NH}$ to the Hermitian part of the Hamiltonian. The full
Hamiltonian $H$ then reads $H=H_0+ H_{int}+H_{NH}$, where
\cite{Fan98}
\begin{subequations}
\begin{eqnarray}
H_0 & = & \omega_a |a  \rangle \langle a|+\sum_{\mu} \int d \omega_{\mu} \omega_{\mu} | \omega_{\mu} \rangle \langle \omega_{\mu}|, \\
H_{int} & = & \lambda \sum_{\mu}  \int d \omega_{\mu}
 \left[ \kappa_{\mu}(\omega_{\mu}) |\omega_{\mu} \rangle \langle a | +
h.c. \right], \\
H_{NH}& = & i(g-\gamma_i)|a  \rangle \langle a|,
\end{eqnarray}
\end{subequations}
with $\langle a| a \rangle=1$, $\langle \omega_{\mu}|
\omega^{'}_{\mu^{'}} \rangle=\delta_{\mu, \mu^{'}}
\delta(\omega_{\mu}-\omega^{'}_{\mu})$, $\langle a | \omega_{\mu}
\rangle=0$, and $\hbar=1$. The coefficients
$\kappa_{\mu}(\omega_{\mu})$ describe the direct coupling between
the localized mode $|a\rangle$ of the microcavity and the
propagating modes $|\omega_{\mu}\rangle$ in the continuum, whereas
$\lambda$ is a dimensionless parameter that measures the strength
of interaction ($\lambda \rightarrow 0$ for a vanishing
interaction). If we write the state $|\psi\rangle$ as
\begin{equation}
|\psi\rangle=c_a(t)|a \rangle+ \sum_{\mu} \int d \omega_{\mu}
c_{\mu}(\omega_{\mu},t) | \omega_{\mu}\rangle
\end{equation}
the following coupled-mode equations for the coefficients $c_a(t)$
and  $c_{\mu}(\omega_{\mu},t)$ are readily obtained from the
equation $ i \partial |\psi\rangle / \partial t=H |\psi\rangle$:
\begin{widetext}
\begin{subequations}
\begin{eqnarray}
i \dot c_a(t) & = & (\omega_a+ig-i\gamma_i)c_a(t)+ \lambda
\sum_{\mu} \int d \omega_{\mu}
\kappa_{\mu}^{*}(\omega_{\mu})c_{\mu}(\omega_{\mu},t)
, \label{cme1}\\
i \dot c_{\mu}(\omega_{\mu},t) & = & \omega_{\mu}
c_{\mu}(\omega_{\mu},t)+\lambda \kappa_{\mu}(\omega_{\mu})c_a(t),
\label{cme2}
\end{eqnarray}
\end{subequations}
\end{widetext}
where the dot stands for the derivative with respect to time $t$.
Note that the power of the microcavity mode is given by
$|c_a(t)|^2$, whereas the total power of the field (cavity plus
structured reservoir) is given by $P(t)=|c_a(t)|^2+\sum_{\mu}\int
d \omega_{\mu} |c_{\mu}(\omega_{\mu},t)|^2$. The threshold
condition for lasing is obtained when an initial perturbation in
the system does not decay with time. From Eqs.(\ref{cme1}) and
(\ref{cme2}) the following power-balance equation can be derived
\begin{equation}
\frac{dP}{dt}=(g-\gamma_i)|c_a|^2, \label{power}
\end{equation}
from which we see that $|c_a|^2 \rightarrow 0$ for any
$g<\gamma_i$, so that the threshold $g=g_{th}$ for laser
oscillation satisfies the condition $ g_{th} \geq \gamma_i$, as
expected.

\subsection{Weak coupling limit: Markovian dynamics}

The temporal evolution of the microcavity-mode amplitude $c_a(t)$
and the condition for laser oscillation can be rigorously obtained
by solving the coupled-mode equations (\ref{cme1}) and
(\ref{cme2}) by means of a Laplace transform analysis, which will
be done in the next section. Here we show that, in the weak
coupling regime ($\lambda \rightarrow 0$) and for a broad band of
continuous modes, coupling of the cavity mode with the neighboring
waveguides leads to the usual Weisskopf-Wigner (exponential)
decay. Though this is a rather standard result (see, e.g.
\cite{Tannoudji}) and earlier derived for a standard Fabry-Perot
laser resonator in Ref.\cite{Lang73} using a Fano diagonalization
technique, for the sake of completeness it is briefly reviewed
here within the model described in Sec.II.A. If the system is
initially prepared in state $|a\rangle$, i.e. if at initial time
$t=0$ there is no field in the neighboring waveguides and $c_a(0)
\neq 0$, an integro-differential equation describing the temporal
evolution of cavity mode amplitude $c_a(t)$ at successive times
can be derived after elimination of the reservoir degrees of
freedom. A formal integration of Eqs.(\ref{cme2}) with initial
condition $c_{\mu}(\omega_{\mu},0)=0$ yields
\begin{equation}
c_{\mu}(\omega_{\mu},t)=-i \lambda \kappa_{\mu}(\omega_{\mu})
\int_{0}^{t} dt' c_a(t') \exp[-i \omega_{\mu}(t-t')].
\label{eliminac}
\end{equation}
After setting $c_a(t)=A(t) \exp(-i \omega_a t)$, substitution of
Eq.(\ref{eliminac}) into Eq.(\ref{cme1}) yields the following {\it
exact} integro-differential equation for the mode amplitude $A(t)$
\begin{equation}
\dot A=(g-\gamma_i)A-\int_{0}^t d \tau G(\tau) A(t-\tau),
\label{integrodiff}
\end{equation}
where $G(\tau)$ is the reservoir response (memory) function, given
by
\begin{equation}
G(\tau)=\lambda^2 \sum_{\mu} \int d \omega_{\mu}
|\kappa_{\mu}(\omega_{\mu})|^2 \exp[-i(\omega_{\mu}-\omega_a)
\tau]. \label{memory}
\end{equation}
Equation (\ref{integrodiff}) clearly shows that the dynamics is
not a Markovian process since the evolution of the mode amplitude
at time $t$ depends  on previous states of the reservoir.
Nevertheless, if the characteristic memory time $\tau_m$ is short
enough (i.e., the spectral coupling coefficients $\kappa_{\mu}$
broad enough) and the coupling weak enough such that $|\dot A / A|
\tau_m \ll 1$, we may replace Eq.(\ref{integrodiff}) with the
following approximate equation
\begin{equation}
\dot A \simeq (g-\gamma_i)A-A(t) \int_{0}^t d \tau G(\tau)  \simeq
(g-\gamma_i)A-(\gamma_R+i\Delta_R)A, \label{Markovian}
\end{equation}
where
\begin{equation}
(\gamma_R+i\Delta_R) = \int_{0}^{t}d \tau G(\tau)
\end{equation}
for $t \gg \tau_m$. In this limit, the dynamics is therefore
Markovian and the reservoir is simply accounted for by a decay
rate $\gamma_R$ and a frequency shift $\Delta_R$. Using the
relation
\begin{equation}
\lim_{t \rightarrow \infty} \int_{0}^t d \tau \exp(-i \omega
\tau)= \pi \delta(\omega)-i \mathcal{P} \left( \frac{1}{\omega}
\right),
\end{equation}
from Eq.(\ref{memory}) the following expressions for the decay
rate $\gamma_R$ and the frequency shift $\Delta_R$ can be derived
\begin{eqnarray}
\gamma_R & = &  \pi \lambda^2 \sum_{\mu} |\kappa_{\mu}(\omega_a)|^2 , \label{decayrate} \\
\Delta_R & = & \lambda^2 \sum_{\mu} \mathcal{P} \int d
\omega_{\mu} \frac{|\kappa_{\mu}(\omega_{\mu})|^2
}{\omega_a-\omega_{\mu}}. \label{frequencyshift}
\end{eqnarray}
The dynamics of the cavity mode field in the Markovian
approximation is therefore standard: an initial field amplitude in
the cavity will exponentially decay, remain stationary
(delta-function lineshape) or exponentially grow (in the early
stage of lasing) depending on whether $g< \gamma$, $g=\gamma$ or
$g> \gamma$, respectively, where $\gamma=\gamma_i+\gamma_R$ is the
total cavity decay rate. The threshold for laser oscillation is
therefore simply given by $g_{th}=\gamma_i+\gamma_R$, i.e.
\begin{equation}
g_{th}=\gamma_i+\pi \lambda^2 \sum_{\mu}
|\kappa_{\mu}(\omega_a)|^2. \label{thmarkovian}
\end{equation}

\section{Field Dynamics beyond the Markovian Limit: general aspects}
Let us assume that the system is initially prepared in state
$|a\rangle$, i.e. that at initial time $t=0$ there is no field in
the neighboring waveguides [$c_{\mu}(\omega_{\mu},0)=0$] whereas
$c_a(0)=1$. The exact solution for the field amplitude $c_a(t)$ of
the microcavity mode at successive times can be obtained by a
Laplace-Fourier transform of Eqs.(\ref{cme1}) and (\ref{cme2}).
Let us indicate by $\hat{c_a}(s)$ and
$\hat{c_{\mu}}(\omega_{\mu},s)$ the Laplace transforms of $c_a(t)$
and ${c_{\mu}}(\omega_{\mu},t)$, respectively, i.e.
\begin{equation}
\hat{c_a}(s)=\int_{0}^{\infty}dt \; c_a(t) \exp(-st)
\label{Laplace}
\end{equation}
and a similar expression for $\hat{c_{\mu}}(\omega_{\mu},s)$. From
the power balance equation (\ref{power}), one can easily show that
the integral on the right hand side in Eq.(\ref{Laplace})
converges for ${\rm Re}(s)> \eta$, where $\eta=0$ for $g-\gamma_i
\leq 0$ or $\eta=g-\gamma_i$ for $g-\gamma_i>0$. The field
amplitude $c_a(t)$ is then written as the inverse Laplace
transform
\begin{equation}
c_a(t)= \frac{1}{2 \pi i } \int_{{\rm B}} ds \; \hat{c}_a(s)
\exp(st) \label{invLaplace}
\end{equation}
where the Bromwich path ${\rm B}$ is a vertical line ${\rm
Re}(s)={\rm const}> \eta$ in the half-plane of analyticity of the
transform, and $\hat{c}_a(s)$ is readily derived after Laplace
transform of Eqs.(\ref{cme1}) and (\ref{cme2}) and reads
\begin{equation}
\hat{c}_a(s)= \frac{i}{is-\omega_a-ig'-\Sigma(s)}
\label{Laplaceca}
\end{equation}
In Eq.(\ref{Laplaceca}), $g'=g-\gamma_i$ is the effective gain
parameter and $\Sigma(s)$ is the self-energy function, which is
expressed in terms of the form factor
\begin{equation}
\Sigma(s)=\int_{\omega_1}^{\omega_2} d \omega
\frac{\mathcal{D}(\omega)}{is-\omega} \label{selfenergy}
\end{equation}
where $\mathcal{D}(\omega)$ is the reservoir structure function,
defined by
\begin{equation}
\mathcal{D}(\omega)=\lambda^2 \sum_{\mu} |\kappa_{\mu}(\omega)|^2.
\end{equation}
In writing Eq.(\ref{selfenergy}), we assumed that the spectrum of
modes of the waveguides (to which the microcavity is coupled)
shows an upper and lower frequency limits $\omega_1$ and
$\omega_2$. We will also assume that $\mathcal{D}(\omega)$ does
not show gaps, i.e. intervals with $\mathcal{D}=0$, inside the
range $(\omega_1,\omega_2)$. The assumption of a finite spectral
extension for the continuous modes is physically reasonable and is
valid for e.g. PC waveguides or CROW. In addition, in order to
avoid the existence of bound states (or polariton modes) for the
passive microcavity coupled to the structured reservoir, we assume
that $\mathcal{D}(\omega)$ vanishes at the boundary of the band,
precisely we require that $\mathcal{D}(\omega) \sim
(\omega-\omega_{1,2})^{\delta_{1,2}}$ as $\omega \rightarrow
\omega_{1,2}$, with $\delta_{1,2}>0$. This condition, which will
be clarified in Sec.III.A, is a necessary requirement to ensure
that the field amplitude $c_a(t)$ fully decays toward zero for $g'=0$.\\
The temporal evolution of $c_a(t)$ is largely influenced by the
analytic properties of $\hat{c}_a(s)$; in particular the
occurrence of a singularity (pole) at $s=s_{pole}$ with ${\rm
Re}(s_{pole}) \geq 0$ may indicate the onset of an instability,
i.e. a lasing regime. The self-energy function $\Sigma(s)$
[Eq.(\ref{selfenergy})], and hence $\hat{c}_a(s)$, are not defined
on the segment of the imaginary axis $s=-i \omega$ with $\omega_1<
\omega < \omega_2$, $s_{1,2}=-i \omega_{1,2}$ being two branch
points. In fact, using the relation
 \begin{equation}
 \lim_{\rho \rightarrow 0^+} \frac{1}{\omega \pm i \rho}=\mathcal{P}\left( \frac{1}{\omega} \right) \mp i
 \pi \delta(\omega), \label{deltaR}
\end{equation}
from Eq.(\ref{selfenergy}) one has
\begin{equation}
\Sigma(s=-i \omega \pm  0^+)= \Delta(\omega) \mp i \pi
\mathcal{D}(\omega), \label{disco}
\end{equation}
($\omega_1<\omega<\omega_2$), where we have set
\begin{equation}
\Delta(\omega)=\mathcal{P} \int_{\omega_1}^{\omega_2} d \omega'
\frac{\mathcal{D}(\omega')}{\omega-\omega'}. \label{Omshift}
\end{equation}
To further discuss the analytic properties of $\hat{c}_a(s)$ and
hence the temporal dynamics of $c_a(t)$, one should distinguish
the cases of passive ($g'=0$) and active ($g'>0$) microcavities.

\subsection{The passive microcavity}
Let us first consider the case of $g'=0$, i.e. of a passive
microcavity with negligible internal losses. In this case
 the full Hamiltonian is Hermitian ($H_{NH}=0$),  and therefore
the analytic properties of $\hat{c}_a(s)$ and spectrum of
$H=H_0+H_{int}$ are ruled as follows (see, for instance,
\cite{Tannoudji,Gaveau95,Nakazato96,Regola}): (i) The eigenvalues
$\omega$ of $H$ are real-valued and comprise the continuous
spectrum $\omega_1< \omega < \omega_2$ of unbounded modes and up
to two isolated real-valued eigenvalues, outside the continuous
spectrum from either sides, which correspond to possible bound (or
polariton) modes \cite{Gaveau95}; (ii) The isolated eigenvalues
are the poles of $\hat{c}_a(s)$ on the imaginary axis outside the
branch cut $- \omega_2<{\rm Im}(s)<- \omega_1$; (iii)
$\hat{c}_a(s)$ is analytic in the full complex plane, apart from
the branch cut and the two possible poles on the imaginary axis
corresponding to bound modes; (iv) In the absence of bound modes
$c_a(t)$ fully decays toward zero, whereas a limited (or
fractional) decay occurs
in the opposite case.\\
From Eq.(\ref{Laplaceca}), the poles $s=-i \Omega$ of
$\hat{c}_a(s)$ outside the branch cut are found as solutions of
the equation:
\begin{equation}
\Omega-\omega_a=\int_{\omega_1}^{\omega_2} d \omega
\frac{\mathcal{D}(\omega)}{\Omega-\omega},
\end{equation}
i.e. [see Eq.(\ref{Omshift})]:
\begin{equation}
\Omega-\omega_a=\Delta(\Omega) \label{bound}
\end{equation}
\begin{figure}
\includegraphics[scale=0.4]{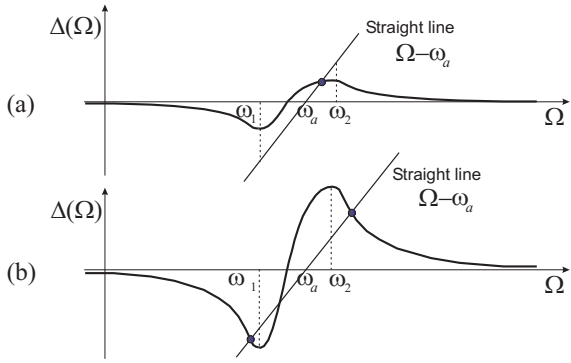} \caption{
 Graphical determination of the roots of Eq.(\ref{bound}) below (a), and above (b) the critical coupling. In (b) the full Hamiltonian $H=H_{0}+H_{int}$ has
 discrete eigenvalues corresponding to bound modes.}
\end{figure}
\noindent with the constraint $\Omega>\omega_2$ or
$\Omega<\omega_1$ \cite{note1}. A graphical solution of
Eq.(\ref{bound}) as intersection of the curves $\Omega-\omega_a$
and $\Delta(\Omega)$ is helpful to decide whether there exist
poles of $\hat{c}_a(s)$, i.e. bound modes (see Fig.1). To this
aim, note that $\Delta(\Omega)>0$ and $d \Delta / d \Omega<0$  for
$\Omega>\omega_2$, $\Delta(\Omega)<0$ and $d \Delta / d \Omega<0$
for $\Omega<\omega_1$, and $\lim_{\Omega \rightarrow \pm \infty}
\Delta(\Omega)=0^{\pm}$. Therefore, Eq.(\ref{bound}) does not have
solutions outside the interval $(\omega_1,\omega_2)$ provided that
$\Delta(\omega_2)<\omega_2-\omega_a$ and
$\Delta(\omega_1)>\omega_1-\omega_a$ [Fig.1(a)]. Such conditions
require at least that $\omega_a$ be internal to the band
$(\omega_1,\omega_2)$, i.e. that the resonance frequency
$\omega_a$ of the microcavity be embedded in the continuum of
decay channels, and that $\mathcal{D}(\omega)$ vanishes as a power
law at the boundary $\omega=\omega_1$ and $\omega=\omega_2$, i.e.
that $\mathcal{D}(\omega) \sim
(\omega-\omega_{1,2})^{\delta_{1,2}}$ as $\omega \rightarrow
\omega_{1,2}$ for some positive integers $\delta_1$ and
$\delta_2$. In fact, if $\mathcal{D}(\omega)$ does not vanish as a
power law at these boundaries, one would have $\Delta(\Omega)
\rightarrow \pm \infty$ as $\Omega \rightarrow \omega_{2},
\omega_1$. Even though $\mathcal{D}(\omega)$ vanishes at the
boundaries, as the coupling strength $\lambda$ is increased either
one or both  of the conditions
$\Delta(\omega_2)>\omega_2-\omega_a$  and
$\Delta(\omega_1)<\omega_1-\omega_a$ can be satisfied [Fig.1(b)],
leading to the appearance of either one or two bound states. The
coupling strength at which a bound state starts to appear is
referred to as {\it critical coupling}. Below the critical
coupling [Fig.1(a)], for the passive microcavity $\hat{c}_a(s)$
does not have poles and a complete decay of $c_{a}(t)$ is
attained. However, owing to non-Markovian effects the decay
dynamics may greatly deviate from the usual Weisskop-Wigner
exponential decay. The exact decay law for $c_a(t)$ is obtained by
the inverse Laplace transform Eq.(\ref{invLaplace}), which can be
evaluated by the residue method after suitably closing the
Bromwich path ${\rm B}$ with a contour in the ${\rm Re}(s)<0$
half-plane (see, e.g. \cite{Tannoudji} pp.220-221, and
\cite{Nakazato96,Regola}). Since the closure crosses the branch
cut $-\omega_2<{\rm Im}(s)< - \omega_1$ on the imaginary axis, the
contour must necessarily pass into the second Riemannian sheet in
the section of the half-plane with  $-\omega_2<{\rm Im}(s)< -
\omega_1$, whereas it remains in the first Riemannian sheet in the
other two sections ${\rm Im}(s)>-\omega_1$ and ${\rm
Im}(s)<-\omega_2$ of the ${\rm Re}(s)<0$ half-plane. To properly
close the contour, it is thus necessary to go back and turn around
the two branch points of the cut at $s=-i \omega_1$ and $s=-i
\omega_2$, following the Hankel paths $h_1$ and $h_2$ as shown in
Fig.2. Note that, while $\hat{c}_a(s)$ is analytic in the first
Riemannian sheet for ${\rm Re}(s)<0$, the analytic continuation
$\hat{c}_{a}^{II}(s)$ of $\hat{c}_a(s)$ from the right [${\rm
Re}(s)>0$] to the left [${\rm Re}(s)<0$] half-plane across the cut
has usually a simple pole at $s=s_{p}$ with ${\rm Re}(s_{p})<0$
and $-\omega_2<{\rm Im}(s_{p})<-\omega_1$ (see Fig.2). Since
$\hat{c}_{a}^{II}(s)=i/[is-\omega_a-\Sigma^{II}(s)]$ with
$\Sigma^{II}(s)=\Sigma(s)- 2 \pi i \mathcal{D}(is)$ [see
Eq.(\ref{disco})], the pole $s_{p}$ is found as a solution of the
equation
\begin{equation}
i s_{p}-\omega_a-\Sigma(s_{p})+2 \pi i \mathcal{D}(i s_{p})=0,
\end{equation}
\begin{figure}
\includegraphics[scale=0.5]{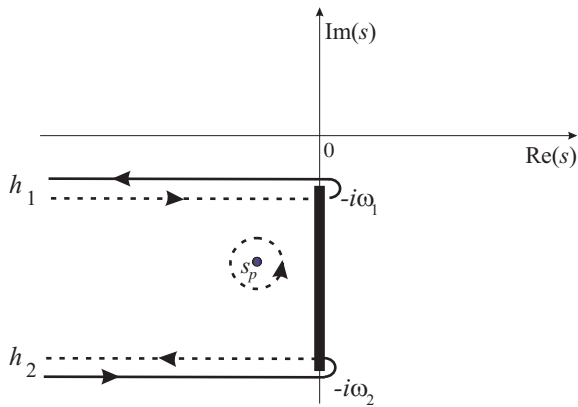} \caption{
Integration contour used to calculate the inverse Laplace
transform of $\hat{c}_a(s)$. The bold solid line on the imaginary
axis is the branch cut. The integration along the solid (dashed)
curves is made on the first (second) Riemannian sheet of
$\hat{c}_a(s)$. $s_p$ is the pole of $\hat{c}_a(s)$ on the second
Riemannian sheet in the ${\rm Re}(s)<0$ half-plane.}
\end{figure}
\noindent i.e.
\begin{eqnarray}
-i \gamma_p+\Delta_p-\int_{\omega_1}^{\omega_2} d \omega
\frac{\mathcal{D}(\omega)}{\omega_a+\Delta_p-i\gamma_p-\omega}+
\nonumber \\
+2 \pi i \mathcal{D}(\omega_a+\Delta_p-i\gamma_p)=0 \label{poloP}
\end{eqnarray}
where we have set
\begin{equation}
s_{p} \equiv - \gamma_{p}-i \omega_a-i\Delta_p.
\end{equation}
After inversion, we then find for $c_a(t)$ the following decay law
\begin{equation}
c_a(t)=\mathcal{Z} \exp[-\gamma_p t
-i(\omega_a+\Delta_p)t]+\mathcal{C}(t), \label{decaylaw}
\end{equation}
where $\mathcal{Z}$ is the residue of $\hat{c}_{a}^{II}(s)$ at the
pole $s_p$, and $\mathcal{C}(t)$ is the contribution from the
contour integration along the Hankel paths $h_1$ and $h_2$ (see
Fig.2):
\begin{eqnarray}
\mathcal{C}(t) & = & \frac{1}{2 \pi i} \int_{s=-\infty-i
\omega_1}^{s=0-i \omega_1} ds \left[
\hat{c}_{a}^{II}(s)-\hat{c}_a(s) \right] \exp(st) + \nonumber \\
 & - & \frac{1}{2 \pi i} \int_{s=-\infty-i \omega_2}^{s=0-i
\omega_2} ds \left[ \hat{c}_{a}^{II}(s) -\hat{c}_a(s) \right]
\exp(st).
\end{eqnarray}
The cut contribution $\mathcal{C}(t)$ is responsible for the
appearance of non-exponential features in the decay dynamics,
especially at short and long times; for an extensive and detailed
analysis we refer the reader to e.g.
Refs.\cite{Nakazato96,Regola}; examples of non-exponential decays
will be presented in Sec.IV.  We just mention here that, in the
weak coupling limit ($\mathcal{D} \rightarrow 0$), from
Eq.(\ref{poloP}) one has that $\gamma_p$ and $\Delta_p$ are small,
and thus using Eq.(\ref{deltaR}) we can cast Eq.(\ref{poloP}) in
the form
\begin{equation}
-i \gamma_p+\Delta_p-\mathcal{P} \int_{\omega_1}^{\omega_2} d
\omega \frac{\mathcal{D}(\omega)}{\omega_a-\omega}+ \pi i
\mathcal{D}(\omega_a) \simeq 0
\end{equation}
from which we recover for the decay rate $\gamma_p$ and frequency
shift $\Delta_p$ of the resonance the same expressions $\gamma_R$
and $\Delta_R$ as given by Eqs.(\ref{decayrate}) and
(\ref{frequencyshift}) in the framework of the Weisskopf-Wigner
analysis. In the strong coupling regime, close to the boundary of
appearance of bound modes, the decay strongly deviates from an
exponential law at any time scale, with the appearance of typical
damped Rabi oscillations (see e.g. Ref. \cite{Tannoudji}, pp.
249-255).

\subsection{Microcavity with gain: lasing condition}

Let us now consider the case of a microcavity with gain, i.e.
$g'>0$. In this case, one (or more) poles $s_p$ of $\hat{c}_a(s)$
on the first Riemannian sheet with ${\rm Re}(s) \geq 0$ may appear
as the modal gain $g'$ is increased, so that the mode amplitude
$c_a(t)$ will grow with time, indicating the onset of an
instability. In this case, the Bromwich path ${\rm B}$ should be
closed taking into account the existence of one (or more than one)
pole in the ${\rm Re}(s) \geq 0$ plane, as shown in Fig.3. For the
case of a simple pole $s_p=-\gamma_p-i\omega_a-i\Delta_p$, the
expression (\ref{decaylaw}) for the temporal evolution of $c_a(t)$
is therefore still valid, where now $\gamma_{p} \leq 0$ and
$\Delta_p$ are found as a solution of the equation [compare with
Eq.(\ref{poloP})]
\begin{equation}
-i \gamma_p-ig'+\Delta_p-\int_{\omega_1}^{\omega_2} d \omega
\frac{\mathcal{D}(\omega)}{\omega_a+\Delta_p-i\gamma_p-\omega}=0.
\label{pologain}
\end{equation}
\begin{figure}
\includegraphics[scale=0.4]{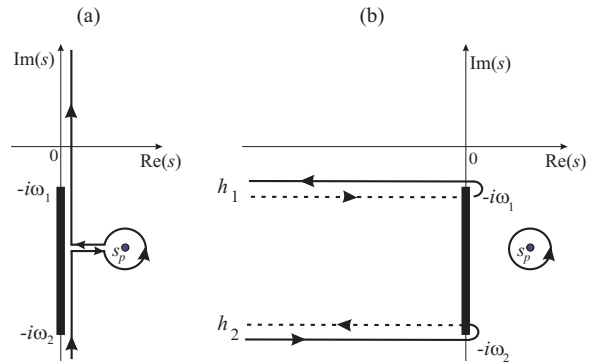} \caption{
(a) Deformation of the Bromwich path for inverse Laplace
transformation with one pole $s_p$ on the ${\rm Re}(s)>0$
half-plane (unstable state). (b) Corresponding integration contour
used to calculate the inverse Laplace transform. The integration
along the solid (dashed) curves is made on the first (second)
Riemannian sheet of $\hat{c}_a(s)$.}
\end{figure}
\noindent As a rather general rule, it turns out that, as $g'$ is
increased, the pole $s_p$ of $\hat{c}_{a}^{II}(s)$, which at
$g'=0$ lies in the ${\rm Re}(s)<0$ plane, crosses the imaginary
axis in the cut region. This crossing changes the decay of
$c_a(t)$ into a non-decaying or growing behavior, and thus it can
be assumed as the threshold for laser oscillation. The modal gain
at threshold, $g^{'}_{th}$, is thus obtained from
Eq.(\ref{pologain}) by setting $\gamma_p=0^-$, i.e.
\begin{equation}
-ig^{'}_{th}+\Delta_p-\Delta(\omega_a+\Delta_p)+ i \pi
\mathcal{D}(\omega_a+\Delta_p)=0, \label{poloth}
\end{equation}
where we used Eq.(\ref{Omshift}) and the relation
\begin{eqnarray}
\int_{\omega_1}^{\omega_2} d \omega
\frac{\mathcal{D}(\omega)}{\omega_a+\Delta_p+i0^+ -\omega}= \\
=\mathcal{P}\int_{\omega_1}^{\omega_2} d \omega
\frac{\mathcal{D}(\omega)}{\omega_a+\Delta_p -\omega}-i \pi
\mathcal{D}(\omega_a+\Delta_p).
\end{eqnarray}
Therefore the threshold for laser oscillation is given by
\begin{equation}
g_{th}=\gamma_i+\pi \mathcal{D}(\omega_a+\Delta_p),
\label{thgeneral}
\end{equation}
where $\Delta_p$ (the frequency shift of the oscillating mode from
the microcavity resonance frequency $\omega_a$) is implicitly
defined by the equation
\begin{equation}
\Delta_p=\mathcal{P}\int_{\omega_1}^{\omega_2} d \omega
\frac{\mathcal{D}(\omega)}{\omega_a+\Delta_p -\omega},
\label{shiftX}
\end{equation}
i.e. $\Omega_{osc}-\omega_a=\Delta (\Omega_{osc})$ with
$\Omega_{osc}=\omega_a+\Delta_p$. It should be noted that, under
the conditions stated in Sec.III.A ensuring that for the passive
microcavity no bound modes exist, Eq.(\ref{shiftX}) admits of (at
least) one solution for $\omega_a+\Delta_p$ inside the range
$(\omega_1,\omega_2)$. The simplest proof thereof can be done
graphically [see Fig.1(a)] after observing that
$\omega_2-\omega_a>\Delta(\omega_2)$ and
$\omega_1-\omega_a<\Delta(\omega_1)$.\\
 The rather simple Eq.(\ref{thgeneral}) provides a generalization of
Eq.(\ref{thmarkovian}) for the laser threshold of the active
microcavity beyond the Markovian approximation and reduces to it
in the limit $\Delta_p \simeq 0$. The frequency shift $\Delta_p$,
however, can not be in general neglected and may strongly affect
the value of $g_{th}$ in the strong coupling regime. In fact, for
a small coupling of the microcavity with the structured reservoir
($\lambda \rightarrow 0$), the shift $\Delta_p$ can be neglected
and therefore $g_{th}$ increases with $\lambda$ according to
Eq.(\ref{thmarkovian}).  However, as $\lambda$ is further
increased up to the critical coupling condition, the shift
$\Delta_p$ is no more negligible, and the oscillation frequency
$\Omega_{osc}=\omega_a+\Delta_p$ at lasing threshold is pushed
toward the boundaries $\omega_1$ or $\omega_2$, where
$\mathcal{D}(\omega)$ and thus $g{'}_{th}$ vanish. In fact, as
$\lambda$ is increased to reach the minimum value between
$\lambda_{I,II}$ defined by the relation \cite{note2}:
\begin{equation}
\lambda_{I,II}^2= (\omega_{1,2}-\omega_a) \left[ \mathcal{P}
\int_{\omega_1}^{\omega_2}d \omega
\frac{\sum_{\mu}|\kappa_{\mu}(\omega)|^2}{\omega_{1,2}-\omega}
\right]^{-1},
\end{equation}
one has $\Omega_{osc} \rightarrow \omega_{1,2}$, and hence $g_{th}
\rightarrow \gamma_i$. Therefore, as $g_{th}$ initially increases
from $\gamma_i$ as the coupling strength is increased from
$\lambda=0$, it must reach a maximum value and then start to
decrease until reaching again the $\gamma_i$ value as $\lambda$
approaches the critical value ($\lambda_{I}$ or $\lambda_{II}$).
As the increase of $g_{th}$ with $\lambda$ in the weak coupling
regime is simply understood as due to the acceleration of the
decay of the microcavity mode into the neighboring waveguides, the
successive decreasing of $g_{th}$ is related to the appearance of
a back-coupling of the field from the continuum (waveguides) into
the microcavity mode, until a bound state is formed at the
critical coupling
strength.\\
As a final remark, it should be noted that the precise dynamical
features and the kind of instability at lasing threshold may
depend on the specific structure function $\mathcal{D}(\omega)$ of
the reservoir. In particular, anomalous dynamical features may
occur at the critical coupling regime, as it will be shown in the
next section.

\section{An exactly-solvable model: the coupling of a microcavity with a coupled resonator
optical waveguide}

To clarify the general results obtained in the previous section,
we present an illustrative example of exactly-solvable model in
which a single-mode and high-$Q$ microcavity is tunneling-coupled
to a CROW structure \cite{Stefanou98,Yariv99,Ozbay00,Olivier01},
which provides the non-markovian decay channel of the microcavity.
In a CROW structure, photons tunnel from one evanescent defect
mode of a cavity to the neighboring one due to overlapping between
the tightly confined modes at each defect site, and therefore
memory effects are expected to be non-negligible whenever the
coupling rate of the microcavity with the CROW becomes comparable
with the CROW hopping rate.

\subsection{The model}
The schematic model of a microcavity tunneling-coupled to a CROW
is shown in Fig.4 for two typical configurations. The CROW
consists of a chain of equally-spaced optical waveguides
\cite{Stefanou98,Yariv99,Ozbay00,Olivier01}, supporting a single
band of propagating modes, and the microcavity is
tunneling-coupled to either one [Fig.4(a)] or two [Fig.4(b)]
cavities of the CROW. For the sake of definiteness, we will
consider the coupling geometry shown in Fig.4(b), though similar
results are obtained for the single-coupling configuration of
Fig.4(a).\\
\begin{figure}
\includegraphics[scale=0.55]{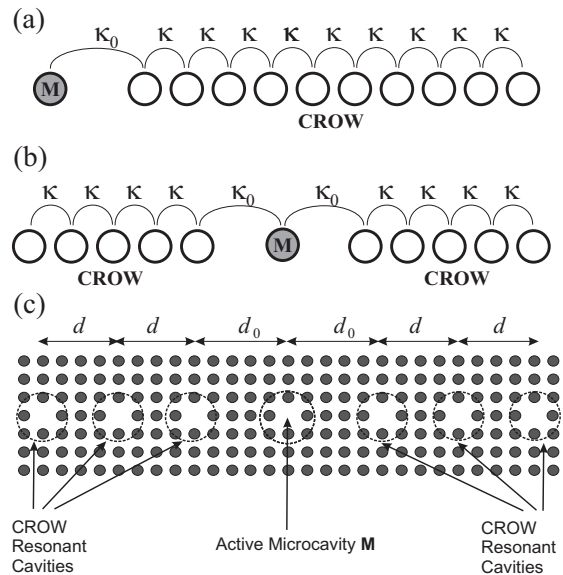} \caption{
Schematic of a microcavity (M) tunneling-coupled to either one (a)
or two (b) cavities of a coupled-resonator optical waveguide. Plot
(c) shows a schematic of a microcavity coupled with a CROW in the
configuration (b) realized on a PC planform made of a square
lattice of air holes with a one-dimensional chain of defects
patterned along the lattice (Ref.\cite{Liu05}).}
\end{figure}
\noindent The microcavity and the CROW can be realized on a same
PC planform (see, e.g., \cite{Liu05,Yanik04}): the CROW is simply
obtained by a one-dimensional periodic array of defects, placed at
distance $d$ and patterned along the lattice to form resonant
cavities with high-$Q$ factors. The microcavity is realized by one
defect in the array, say the one corresponding to index $n=0$,
which can have a resonance frequency $\omega_a$ different from
that of adjacent defects and placed at a larger distance $d_0 \geq
d$ than the other cavities [see Fig.4(c)]. The CROW supports a
continuous band of propagating modes whose dispersion relation, in
the tight-binding approximation, is given by \cite{Yariv99}
\begin{equation}
\omega(k)=\omega_0-2 \kappa \cos(kd),
\end{equation}
where $\kappa$ is the hopping amplitude between two consecutive
cavities of the CROW, $d$ is the length of the unit cell of the
CROW, $k$ is the Bloch wave number, and $\omega_0$ is the central
frequency of the band. The resonance frequency $\omega_m$ of the
microcavity is assumed to be internal to the CROW band, i.e.
 $\omega_0-2 \kappa<\omega_m<\omega_0+2 \kappa$. The microcavity is
tunneling-coupled to the two adjacent cavities of the CROW, and we
denote by $\kappa_0$ the hopping amplitude. The ratio $\kappa_0 /
\kappa$ and the position of $\omega_m$ inside the CROW band can be
properly controlled by changing the geometrical parameters of the
defects and the ratio $d_0/d$. In particular, in the limiting case
where the microcavity has the same geometry and distance of the
other CROW cavities, one has $\kappa_0=\kappa$ and
$\omega_m=\omega_0$. An excellent and simple description of light
transport in the system is provided by a set of coupled-mode
equations for the amplitudes $a_n$ of modes in the cavities (see,
e.g., \cite{Yariv99,Yanik04})
\begin{subequations}
\begin{eqnarray}
i\dot a_n & = & -\kappa(a_{n+1}+a_{n-1}) \; \; (|n| \geq 2) \\
i\dot a_{-1} & = & -\kappa a_{-2}- \kappa_0 c_a \\
i\dot c_{a} & = & -\kappa_0 (a_{-1}+ a_1)+(\omega_a+ig) c_a \\
i \dot a_{1} & = & -\kappa a_{2}-\kappa_0 c_a
\end{eqnarray}
\end{subequations}
where $c_a$ is the amplitude of the microcavity mode, $g$ is its
effective modal gain per unit time, and
$\omega_a=\omega_m-\omega_0$ is the frequency detuning between the
microcavity resonance frequency $\omega_m$ and the central
frequency $\omega_0$ of the CROW band. For e.g. a CROW built in a
GaAs-based PC with a square lattice of air holes in the design of
Ref.\cite{Liu05}, a typical value of the cavity coupling
coefficient turns out to be $\kappa \simeq 700-800$ GHz and
$\omega_0 / \kappa \sim 3 \times 10^3$ at the $\lambda_0 =850$ nm
operation wavelength. Note that in writing Eqs.(38), we have
neglected the internal losses of the CROW cavities; a reasonable
value of the $Q$-factor for a realistic microcavity is
$Q=\omega_0/(2 \gamma_{loss}) \sim  10^6$ \cite{Armani03}, which
would correspond to a cavity loss rate $\gamma_{loss} \sim 1 $ GHz
to be added in Eqs.(38). This loss rate, however, is about
two-to-three orders of magnitude smaller than the cavity coupling
coefficient $\kappa$, and therefore on a short time scale
non-Markovian dynamical effects should be observed even in
presence of CROW losses. The effects of reservoir (CROW) losses
will be briefly discussed at the end of the section.\\
To study the temporal evolution of an initial field in the
microcavity, Eqs.(38) are solved with the initial condition
$a_n(0)=0$ and $c_a(0)=1$. An integral representation for the
solution of Eqs.(38) might be directly derived in the time domain
by an extension of the technique described in
Refs.\cite{Longhi06a,Longhi06b}, where a system of coupled-mode
equations similar to Eqs.(38), but in the conservative (i.e.
$g=0$) case, was considered. However, we prefer here to formally
place Eqs.(38) into the more general Hamiltonian formalism of
Sec.II and then use the Laplace transform analysis developed in
the previous section to obtain the temporal evolution for
$c_a(t)$. To this aim, in Appendix we prove that $c_a(t)$ may be
obtained as a solution of the following equations, which have the
canonical form (3) with a simple continuum of modes acting as a
decay channel
\begin{subequations}
\begin{eqnarray}
i \dot c_a(t) & = & (\omega_a+ig)c_a+ \lambda
\int_{-2 \kappa}^{ 2 \kappa} d \omega  \kappa_{\mu}(\omega) c(\omega,t) \label{cme1a}\\
i \dot c(\omega,t) & = & \omega c(\omega,t)+\lambda
\kappa_{\mu}(\omega) c_a(t) \label{cme2a}
\end{eqnarray}
\end{subequations}
with
\begin{equation}
\lambda \kappa_{\mu}(\omega)=\kappa_0 \sqrt{\frac{2}{\pi \kappa}}
\left[1-\left( \frac{\omega}{2 \kappa} \right)^2 \right]^{1/4}.
\end{equation}
Note that the reservoir structure function for this model, defined
for $\omega_1<\omega<\omega_2$ with $\omega_1=-2 \kappa$ and
$\omega_2=2 \kappa$, is simply given by
\begin{equation}
\mathcal{D}(\omega)=\frac{2 \kappa_{0}^2}{\pi \kappa}
\sqrt{1-\left( \frac{\omega}{2 \kappa} \right)^2}. \label{sfCROW}
\end{equation}
With this reservoir structure function, the self-energy
[Eq.(\ref{selfenergy})] can be calculated in an exact way and
reads
\begin{equation}
\Sigma(s)=i \left( \frac{\kappa_0}{\kappa} \right)^2 \left[
s-\sqrt{4 \kappa^2+s^2} \right].
\end{equation}
The function $\Delta(\omega)$, as defined by Eq.(\ref{Omshift}),
then reads
\begin{equation}
\Delta(\omega)= \left\{
\begin{array}{lr}
(\kappa_0 / \kappa)^2 \omega & |\omega| < 2 \kappa \\
(\kappa_0 / \kappa)^2 \left[\omega-\sqrt{\omega^2-4 \kappa^2}
\right] & \omega > 2 \kappa \\
(\kappa_0 / \kappa)^2 \left[\omega+\sqrt{\omega^2-4 \kappa^2}
\right] & \omega < -2 \kappa
\end{array}
\right. \label{shiftCROW}
\end{equation}
Note that the coupling strength between the microcavity and the
CROW is determined by the ratio $\kappa_0 / \kappa$, the limit
$\kappa_0 / \kappa \rightarrow 0$ corresponding to the weak
coupling regime.

\subsection{The passive microcavity: from exponential decay to damped Rabi oscillations}

Let us consider first the case $g=0$. The conditions for the
non-existence of bound modes, i.e. for a complete decay of
$c_a(t)$, are $\omega_2-\omega_a \geq \Delta(\omega_2)$ and
$\omega_1-\omega_a \leq \Delta(\omega_1)$ (see Sec.III.A), which
using Eq.(\ref{shiftCROW}) read explicitly
\begin{equation}
\left( \frac{\kappa_0}{\kappa} \right)^2 -1  \leq
\frac{\omega_a}{2 \kappa} \leq 1-\left( \frac{\kappa_0}{\kappa}
\right)^2.
\end{equation}
Note that, as a necessary condition, this relation implies that
$|\omega_a| \leq 2 \kappa$ and $(\kappa_0 / \kappa)^2 \leq 1$.
Note also that the the critical coupling regime is reached at
$(\kappa_0 / \kappa)=\sqrt{1 -|\omega_a|/(2 \kappa)}$. For a
coupling strength $(\kappa_0 / \kappa)$ above such a value, the
decay of $c_a(t)$ is imperfect due to the existence of bound
modes between the microcavity and the CROW; this case will not be considered here further.\\
The temporal decay law for the mode amplitude $c_a(t)$ can be
generally expressed using the general relation (\ref{decaylaw}),
which highlights the existence of the exponential
(Weisskopf-Wigner) decaying term plus its correction due to the
contribution of the Hankel paths. Perhaps, for the
microcavity-CROW system it is more suited to make the inverse
Laplace transform on the first Riemannian sheet of $\hat{c}_a(s)$
by closing the Bromwich path ${\rm B}$ with a semicircle with
radius $R \rightarrow \infty$ in the ${\rm Re}(s)<0$ half-plane
after excluding the branch cut from the domain by the contour
$\sigma$ as shown in Fig.5. Since in this case there are no
singularities of $\hat{c}_a(s)$, we simply obtain
\begin{equation}
c_{a}(t)=\frac{1}{2 \pi } \oint_{\sigma} ds \;
\frac{\exp(st)}{is-\omega_a-\Sigma(s)}
\end{equation}
which, using Eq.(\ref{disco}), reads explicitly
\begin{figure}
\includegraphics[scale=0.55]{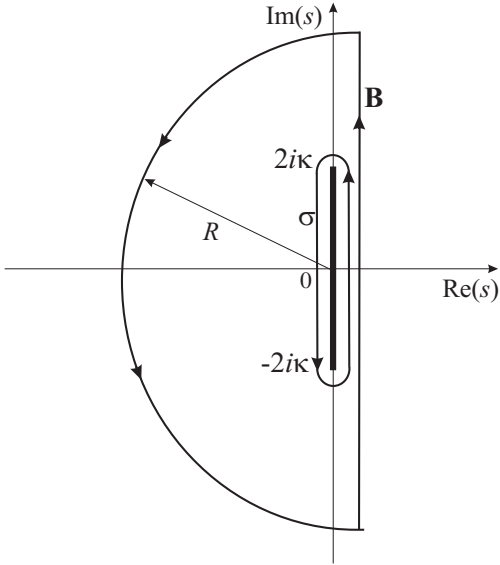} \caption{
Integration contour used for the inverse Laplace transform in the
passive microcavity-CROW system.}
\end{figure}

\begin{eqnarray}
c_{a}(t) & = & \frac{i}{2 \pi } \int_{\omega_1}^{\omega_2} d
\omega \left[ \frac{\exp(-i \omega
t)}{\omega-\omega_a-\Sigma(-i\omega+0^+)}+ \right. \nonumber \\
& - & \left. \frac{\exp(-i \omega
t)}{\omega-\omega_a-\Sigma(-i\omega-0^+)} \right] = \nonumber \\
& = & \int_{\omega_1}^{\omega_2} d \omega
\frac{\mathcal{D}(\omega) \exp(-i \omega
t)}{[\omega-\omega_a-\Delta(\omega)]^2+ \pi^2
\mathcal{D}^2(\omega)}.
\end{eqnarray}
For the microcavity-CROW model, one then obtains
\begin{widetext}
\begin{equation}
c_{a}(t)=\frac{1}{2 \pi} \frac{\kappa_{0}^2}{\kappa^3} \int_{-2
\kappa}^{2 \kappa} d \omega \frac{\exp(-i \omega t)
\sqrt{1-(\omega/2 \kappa)^2}}{\left\{ (\omega/2 \kappa) \left[
1-(\kappa_0 / \kappa)^2 \right]- (\omega_a / 2 \kappa)
\right\}^2+(\kappa_0 / \kappa)^4 [1-\omega^2 / (4 \kappa^2)]}.
\label{intcaCROW}
\end{equation}
\end{widetext}
The integral on the right hand side in Eq.(\ref{intcaCROW}) can be
written in a more convenient form with the change of variable
$\omega=-2 \kappa \cos Q$, yielding
\begin{widetext}
\begin{equation}
c_a(t)=\frac{1}{\pi} \int_{0}^{\pi} dQ \frac{(k_0 / \kappa)^2
\sin^2 Q \exp(2i \kappa t \cos Q)}{\left[ (\omega_a/ 2 \kappa)
+\cos Q-(\kappa_0 / \kappa)^2 \cos Q \right]^2+(\kappa_0 /
\kappa)^4 \sin^2 Q }. \label{intrep}
\end{equation}
\end{widetext}
In this form, the integral can be written \cite{Longhi06a} as a
series of Bessel functions of first kind and of argument $2 \kappa
t$ (Neumann series). Special cases, for which a simple expression
for $c_a(t)$ is available, are those corresponding to $\omega_a=0$
and $\kappa_0=\kappa$, for which
\begin{equation}
c_a(t)=J_0(2 \kappa t),
\end{equation}
and to $\omega_a=0$ and $\kappa_0=\kappa/ \sqrt 2$, for which
\begin{equation}
c_a(t)=\frac{J_1(2 \kappa t)}{\kappa t} .
\end{equation}
Note that the former case corresponds to a critical coupling
regime, where $\hat{c}_a(s)$ has two singularities at $s=\pm 2i
\kappa+0^+$. The residues of $\hat{c}_a(s)$ at these
singularities, however, vanish, and therefore the field $c_a(t)$
fully decays toward zero with an asymptotic power law $\sim
1/t^{1/2}$. In general, an inspection of the singularities of the
$\hat{c}_a(s)$ reveals that, for $\omega_a \neq 0$, at the
critical coupling strength
$(\kappa_0/\kappa)=\sqrt{1-|\omega_a|/(2 \kappa)}$ the Laplace
transform $\hat{c}_a(s)$ has one singularity at either $s_p=2i
\kappa+0^+$ or $s_p=-2i \kappa+0^+$ of type $\hat{c}_a(s) \sim 1/
\sqrt{s-s_p}$.\\
The asymptotic decay behavior of $c_a(t)$ at long times can be
determined by the application of the method of the stationary
phase to Eq.(\ref{intrep}). One then finds that at the critical
coupling the field $c_a(t)$ decays toward zero with an asymptotic
power law $\sim 1/t^{1/2}$, whereas below the critical coupling
the decay is
faster with an asymptotic decay $\sim 1/t^{3/2}$.\\
Typical examples of non-exponential features in the decay process
as the coupling strength is increased are shown in Fig.6 for
$\omega_a=0$. The curves in the figures have been obtained by a
direct numerical solution of Eqs.(38). Note that, as for weak
coupling the exponential (Weisskopf-Wigner) decay law is retrieved
with a good approximation [see Fig.6(a)], as the coupling strength
$\kappa_0 / \kappa$ is increased the decay law strongly deviates
from an exponential behavior. Note in particular the existence of
strong oscillations, which are fully analogous to damped Rabi
oscillations found in the atom-photon interaction context
\cite{Tannoudji}. For $\omega_0 \neq 0$, the oscillatory behavior
of the long-time power-law decay is less pronounced and may even
disappear (see Ref.\cite{Longhi06a}).

\begin{figure}
\includegraphics[scale=0.43]{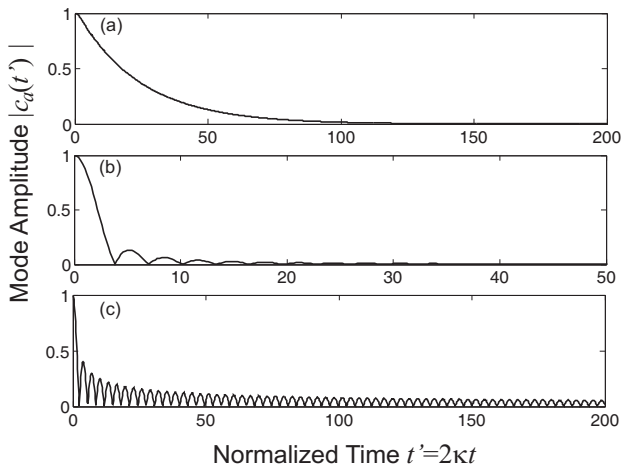} \caption{
Decay of the mode amplitude $|c_a(t)|$ in a passive
microcavity-CROW system for $\omega_a=0$ and for increasing values
of coupling strength: (a) $\kappa_0 / \kappa=0.2$, (b) $\kappa_0 /
\kappa=0.707$, and (c) $\kappa_0 / \kappa=1$ (critical coupling).}
\end{figure}

\subsection{Microcavity with gain}

Let us consider now the case $g \geq 0$. In order to determine the
threshold for laser oscillation, we have to distinguish three
cases depending on the value of the coupling strength $\kappa_0 / \kappa$.\\
\\
{\it (i) Lasing condition below the critical coupling.} In this
case, corresponding to  $ \kappa_0/ \kappa < \sqrt{1-|\omega_a|/(2
\kappa)}$, the threshold for laser oscillation is readily obtained
from Eqs.(\ref{thgeneral}), (\ref{shiftX}), (\ref{sfCROW}) and
(\ref{shiftCROW}). The frequency $\Omega_{osc}$ of the oscillating
mode is given by $\Omega_{osc}=\omega_a /[1-(\kappa_0 /
\kappa)^2]$, and the gain for laser oscillation is thus given by
\begin{equation}
g_{th}=2 \kappa \left( \frac{\kappa_0}{\kappa} \right)^2
\sqrt{1-\left[ \frac{\omega_a/(2 \kappa)}{1-(\kappa_0 / \kappa)^2
}\right]^2}.
\end{equation}
The typical behavior of normalized threshold gain $g_{th}/(2
\kappa)^2$ versus the coupling strength $(\kappa_0 / \kappa)$ is
shown in Fig.7.
\begin{figure}
\includegraphics[scale=0.5]{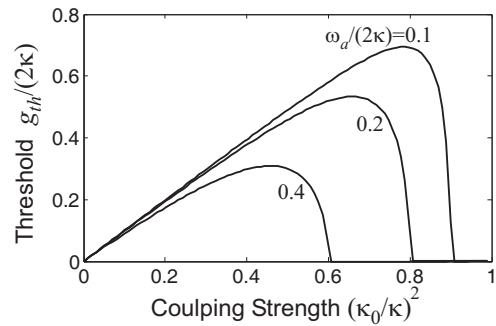} \caption{
Behavior of normalized threshold gain $g_{th}/(2 \kappa)$ versus
the coupling strength $(\kappa_0 / \kappa)^2$ for a few values of
the ratio $\omega_a / (2 \kappa)$.}
\end{figure}
\noindent Note that, according to the general analysis of
Sec.III.B, the threshold for laser oscillation first increases as
the coupling strength is increased, but then it reaches a maximum
and then decreases toward zero as the critical coupling strength
is attained. At $g=g_{th}$, $\hat{c}_a(s)$ has a simple pole at
$s=s_p=-i\Omega_{osc}+0^+$, whereas as $g$ is increased above
$g_{th}$ the pole $s_p$ invades the ${\rm Re}(s)>0$ half-plane.
Therefore, the onset of lasing is characterized by an amplitude
$|c_a(t)|$ which asymptotically decays toward zero for $g<g_{th}$,
reaches a steady-state and nonvanishing value at $g=g_{th}$ (the
field does not decay nor grow asymptotically), whereas it grows
exponentially (in the early lasing stage) for $g>g_{th}$ with a
growth rate $\sigma(g)={\rm Re}(s_p)$ (see Fig.8). This
instability scenario is the usual one encountered in the
semiclassical theory of laser oscillation as a second-order phase
transition \cite{note3}. However, the temporal dynamics at the
onset of lasing shows unusual oscillations [see Fig.8(a)] which
are a signature of non-Markovian dynamics. In addition, as in the
Markovian limit the growth rate $\sigma$ should increase linearly
with $g-g_{th}$, in the strong coupling regime the growth rate
$\sigma$ shows  near threshold an unusual non-linear
behavior, as shown in Fig.8(b).\\
\begin{figure}
\includegraphics[scale=0.42]{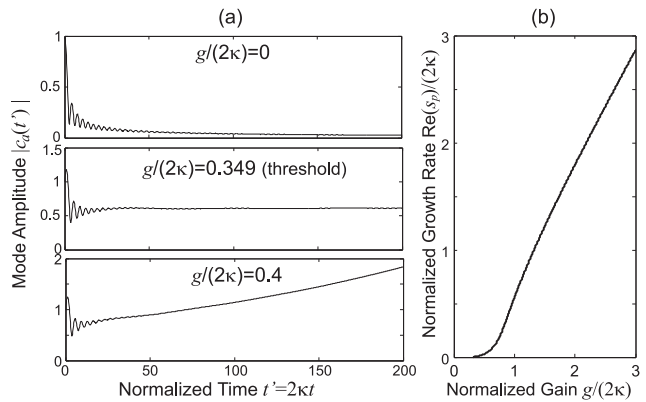} \caption{
(a) Behavior of mode amplitude $|c_a(t')|$ versus normalized time
$t'=2 \kappa t$ for  $(\kappa_0 / \kappa)^2=0.8$, $\omega_a / (2
\kappa)=0.18$, and for increasing values of normalized gain $g/ (2
\kappa)$. (b) Behavior of normalized growth rate versus normalized
gain for $(\kappa_0 / \kappa)^2=0.8$ and $\omega_a / (2
\kappa)=0.18$.}
\end{figure}
\\
{\it (ii) Lasing condition at the critical coupling with $\omega_a
\neq 0$.} A different dynamics occurs when the coupling strength
$\kappa/\kappa_0$ reaches the critical limit $ \kappa_0/
\kappa=\sqrt{1-|\omega_a|/(2 \kappa)}$. As discussed in Sec.IV.B,
at $g=0$ the Laplace transform $\hat{c}_a(s)$ has a singularity at
either $s_p=2 i \kappa$ or $s_p=-2 i \kappa$, however $s_p$ {\it
is not} a simple pole and $c_a(t)$ asymptotically decays toward
zero. For $\omega_a \neq 0$, i.e. for $(\kappa/ \kappa_0)<1$, as
$g$ is increased just above zero $\hat{c}_a(s)$ shows a simple
pole with a growth rate $\sigma={\rm Re}(s_p)>0$ which slowly
increases with $g$ at the early stage, as shown in Fig.9. In the
figure, a typical temporal evolution of $c_a(t)$ is also shown.
Note that in this case {\it there is not} a value of $g$ for which
the field amplitude $c_a(t)$ does not grow nor decay, i.e. the
intermediate situation shown in Fig.8(a) is missed in Fig.9(a):
for $g=0$ the amplitude decays, however for $g=0^+$ it always
grows exponentially. The transition describing the passage of
laser from below to above threshold in the linear stage of the
instability is therefore quite unusual at the critical coupling.\\
\begin{figure}
\includegraphics[scale=0.42]{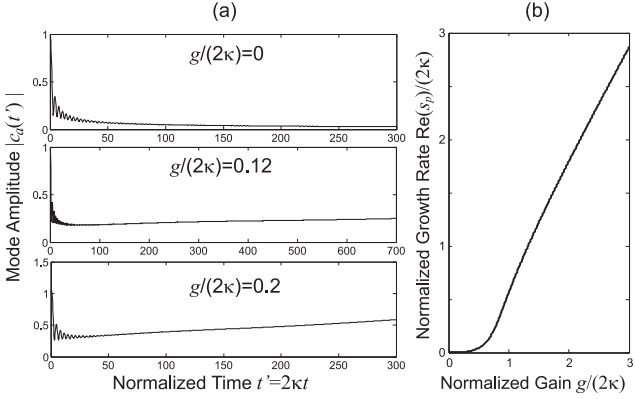} \caption{
Same as Fig.8, but for parameter values $(\kappa_0 /
\kappa)^2=0.8$ and $\omega_a / (2 \kappa)=0.2$ (critical
coupling). Note that in this case there exists no lasing threshold
in the traditional sense.}
\end{figure}
\\
{\it (iii) Lasing condition at the critical coupling with
$\omega_a = 0$.} A somewhat singular behavior occurs at the
critical coupling when $\omega_a=0$, and therefore $\kappa_0 /
\kappa=1$. This case corresponds to consider a periodic CROW in
which one of the cavities is pumped and acts as the microcavity in
our general model. For $\omega_a = 0$ and $\kappa_0 / \kappa=1$,
the Laplace transform $\hat{c}_a(s)$ is explicitly given by
\begin{equation}
\hat{c}_a(s)=\frac{1}{-g+\sqrt{s^2+4 \kappa^2}}.
\end{equation}
To perform the inversion, one needs to distinguish four cases.\\
(a) $g=0$. For $g=0$, the field $c_a(t)$ decays according to
\begin{equation}
c_a(t)=J_0(2 \kappa t)
\end{equation}
as shown in Sec.IV.B.\\
\\
 (b) $0<g<2 \kappa$. In this case
$\hat{c}_a(s)$ has two simple poles on the first Riemannian sheet
at $s_{1,2}=\pm i \sqrt{4 \kappa^2-g^2}+0^+$. The inversion can be
performed by closing the Bromwich path ${\rm B}$ with the contour
shown in Fig.10, where along the dashed curves the integrals are
performed on the second Riemannian sheet. One then obtains
\begin{equation}
c_a(t)=\frac{2 g}{\sqrt{4 \kappa^2-g^2}} \sin \left( \sqrt{4
\kappa^2-g^2 }t \right) +\mathcal{C}(t)
\end{equation}
where the first term on the right hand side in the equation arises
from the residues at poles $s_{1,2}$, whereas $\mathcal{C}(t)$ is
the contribution from the contour integration along the Hankel
paths $h_1$ and $h_2$, which asymptotically decays toward zero as
$t \rightarrow \infty$. Note that, after an initial transient, the
amplitude $|c_a(t)|$ steadily oscillates in time with frequency
$\sqrt{4 \kappa^2-g^2}$ and amplitude $2 g / \sqrt{4 \kappa^2-g^2
}$. Note also that the amplitude and period of oscillations
diverge as the modal gain $g$ approaches $2
\kappa^-$.\\
\begin{figure}
\includegraphics[scale=0.5]{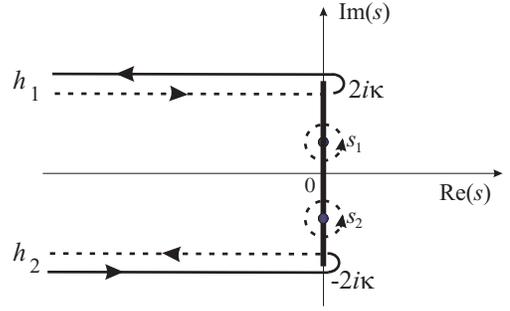} \caption{
Integration contour used to calculate the inverse Laplace
transform for $\omega_a=0$, $\kappa_0 / \kappa=1$ and for $0<g/(2
\kappa) <1$. The integration along the solid (dashed) curves is
made on the first (second) Riemannian sheet of $\hat{c}_a(s)$.
$s_{1,2}$ are the two poles of $\hat{c}_a(s)$ on the imaginary
axis inside the cut.}
\end{figure}
\\
(c) $g=2 \kappa$. In this case, $\hat{c}_a(s)$ has a single pole
of second-order in $s=0^+$, and therefore to perform the inversion
it is worth separating the singular and non-singular parts of $
\hat{c}_a(s)$ as
\begin{equation}
\hat{c}_a(s)= \frac{4 \kappa}{s^2}+f(s)
\end{equation}
where $f(s)$ has no singularities on the imaginary axis. After
inversion one then obtains
\begin{equation}
c_a(t)= 4 \kappa t +\frac{1}{2 \pi } \int_{-\infty}^{\infty} d
\omega \; f(-i \omega+0^+) \exp(-i \omega t), \label{lineargrowth}
\end{equation}
where the second term on the right-hand side in the above equation
asymptotically decays toward zero. Therefore, we may conclude that
at $g=2 \kappa$ the mode amplitude $c_a(t)$ is dominated by a
secular growing term {\it which is not exponential}.\\
\\
(d) $g>2 \kappa$. In this case, $\hat{c}_a(s)$ has an unstable
simple pole at $s_p=(g^2-4 \kappa^2)^{1/2}$, and therefore the
solution $c_a(t)$ grows exponentially with time.\\
\\
The dynamical scenario described above for $\omega_a=0$ and
$\kappa_0 / \kappa=1$  is illustrated in Fig.11. Note that in this
case there is some uncertainty in the definition of laser
threshold, since there exists {\it an entire interval} of modal
gain values, from $g=0^+$ to $g=2 \kappa^-$, at which an initial
field in the cavity does not grow nor decay.\\
\begin{figure}
\includegraphics[scale=0.5]{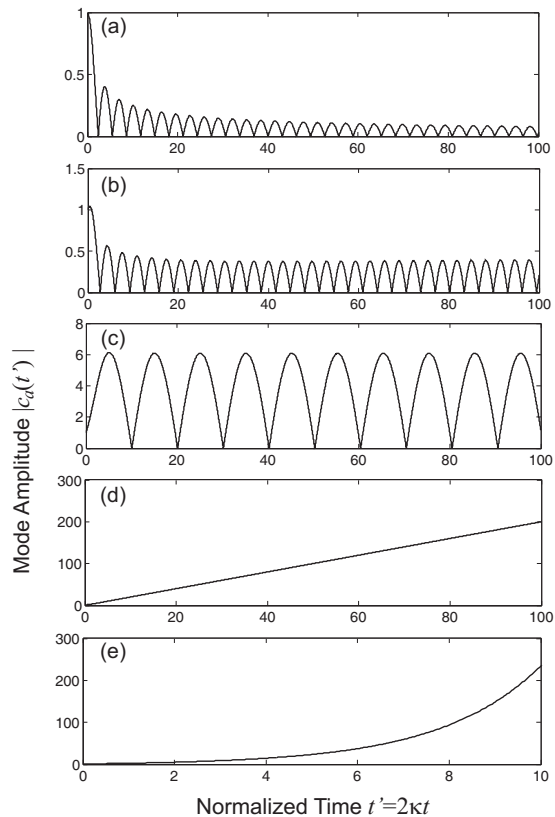} \caption{
Behavior of mode amplitude $c_a(t')$ versus normalized time $t'=2
\kappa t$ for $\omega_a=0$, $\kappa_0 / \kappa=1$ (critical
coupling) and for increasing values of normalized gain: (a)
$g/(2\kappa)=0$, (b) $g/(2\kappa)=0.2$, (c) $g/(2\kappa)=0.95$,
(d) $g/(2\kappa)=1$, and (d) $g/(2\kappa)=1.1$. }
\end{figure}
\\
As a final comment, we briefly discuss the effects of internal
losses of the CROW cavities, which have been so far neglected, on
the temporal evolution of the mode amplitude $c_a(t)$. In the case
where all the cavities in the CROW have the same loss rate
$\gamma_{loss}$, the temporal evolution of $c_a(t)$ is simply
modified by the introduction of an additional exponential damping
factor $\exp(-\gamma_{loss}t)$, i.e. $c_a(t) \rightarrow c_a(t)
\exp(-\gamma_{loss}t)$. This additional decay term would therefore
shift the threshold for laser oscillation to higher values and,
most importantly for our analysis, it might hinder non-Markovian
dynamical effects discussed so far. However, for a small value of
$\gamma_{loss} / \kappa$ (e.g. $\gamma_{loss} / \kappa \sim 0.01$
for the numerical values given in Ref.\cite{Liu05}), non-Markovian
effects should be clearly observable in the transient field
dynamics for times shorter than $\sim 1 /\gamma_{loss}$. As an
example, Fig.12 shows the dynamical evolution of the mode
amplitude $|c_a(t)|$ for the same parameter values of Fig.11,
except for the inclusion of a CROW loss rate $\gamma_{loss}=0.01
\kappa$. It is worth commenting on the dynamical behavior of
Fig.12(d) corresponding to $g=2 \kappa$. In this case, using
Eq.(\ref{lineargrowth}) and disregarding the decaying term on the
right hand side in Eq.(\ref{lineargrowth}), one can write
\begin{equation}
c_a(t) \sim 4 \kappa t \exp(-\gamma_{loss} t).
\end{equation}
Note that in the early transient stage the initial mode amplitude
stored in the microcavity linearly grows as in Fig.11(d), however
it reaches a maximum and then it finally decays owing to the
prevalence of the loss-induced exponential term over the linear
growing term. Therefore, though the microcavity is {\it below}
threshold for oscillation as an initial field in the cavity
asymptotically decays to zero, before decaying an initial field is
subjected to a {\it transient amplification}. The maximum
amplification factor in the transient is about $\sim 2 \kappa /
\gamma_{loss}$, and can be therefore relatively large in high-$Q$
microcavities. Such a transient growth despite the asymptotic
stability of the zero solution should be related to the
circumstance that for $g \neq 0$ the system (38) is non-normal
\cite{note4}: though its eigenvalues have all a negative real
part, the system can sustain a transient energy growth. The
transient amplification shown in Fig.12(d) is therefore analogous
to non-normal energy growth encountered in other hydrodynamic
\cite{Trefethen93,Farrell94,Farrell96} and optical
\cite{Kartner99, Longhi00,Firth05} systems and it is an indicator
of a major sensitivity of the system to noise.
\begin{figure}
\includegraphics[scale=0.5]{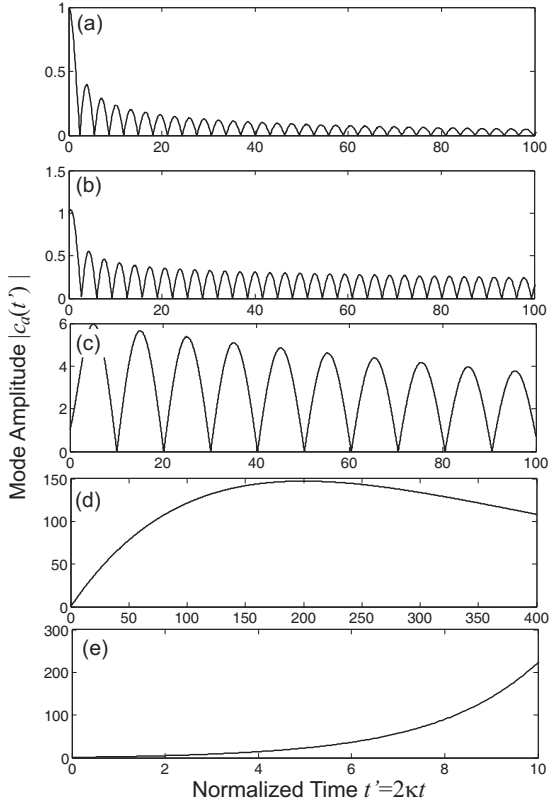} \caption{
Same as Fig.11, but in presence of CROW losses ($\gamma_{loss}/
\kappa=0.01$).}
\end{figure}

\section{Conclusions}

In this work it has been analytically studied, within a rather
general Hamiltonian model [Eqs.(1)], the dynamics of a classical
field in a single-mode optical microcavity coupled to a structured
continuum of modes (reservoir) beyond the usual Weisskopf-Wigner
(Markovian) approximation. Typical non-Markovian effects for the
passive microcavity are non-exponential decay and damped Rabi
oscillations (Sec.III.A). In presence of gain, the general
condition for laser oscillation, that extends the usual gain/loss
rate balance condition of elementary laser theory, has been
derived (Sec.III.B), and the behavior of the laser threshold
versus the microcavity-reservoir coupling has been determined. The
general results have been specialized for an exactly-solvable
model, which can be implemented in a photonic crystal with
defects: an optical microcavity tunneling-coupled to a
coupled-resonator optical waveguide (Sec.IV). A special attention
has been devoted to study the transition describing laser
oscillation at the critical coupling between the cavity and the
waveguide (Sec.IV.C). Unusual dynamical effects, which are a clear
signature of a non-Markovian dynamics, have been illustrated,
including: the existence of a finite interval of modal gain where
the field oscillates without decaying nor growing, the gain
parameter controlling the amplitude and period of the
oscillations; a linear (instead of exponential) growth of the
field at the onset of instability for laser oscillation; and the
existence of transient (non-normal) amplification of the field
below laser threshold when intrinsic losses of the microcavity are
considered. It is envisaged that, though non-Markovian effects are
not relevant in standard laser resonators in which the field
stored in the cavity is coupled to the  broad continuum of modes
of the external open space by a partially-transmitting mirror
\cite{Lang73}, they should be observable when dealing with
high-$Q$ microcavities coupled to waveguides, which act as a
structured decay channel for the field stored in the microcavity.

\appendix
\section{}
In this Appendix it is proved the equivalence between coupled-mode
equations (38) in the tight-binding approximation and the
canonical formulation for the decay of a discrete state into a
continuum provided by Eqs.(39). To this aim, let us first note
that, owing to the inversion-symmetry of the initial condition
$a_{-n}(0)=a_n(0)=0$ ($n \neq 0$), it can be readily shown that
the solution $a_n(t)$ maintains the same symmetry at any time,
i.e. $a_{-n}(t)=a_n(t)$ for $t \geq 0$. Let us then introduce the
 continuous function of the real-valued parameter $Q$
\begin{equation}
\phi(Q,t)=\sum_{n=1}^{\infty} a_n(t) \sin(nQ),
\end{equation}
where $Q$ is taken inside the interval $[0,\pi]$. Using the
relation
\begin{equation}
\int_{0}^{\pi} dQ \; \sin(nQ) \sin(mQ)= \frac{\pi}{2} \delta_{m,n}
\; \; \; (m,n \geq 1)
\end{equation}
the amplitudes $a_n$ of modes in the CROW are related to the
continuous field $\phi$ by the simple relations
\begin{equation}
a_n(t)= \frac{2}{\pi} \int_{0}^{\pi}dQ \; \phi(Q,t) \sin(nQ)
\end{equation}
($n \geq 1$). The equation of motion for $\phi$ is readily
obtained from Eqs.(38) and reads
\begin{equation}
i \frac{\partial \phi}{\partial t}=-2 \kappa \cos(Q) \phi-\kappa_0
\sin(Q) c_a
\end{equation}
whereas the equation for $c_a$, taking into account that
$a_{-1}+a_1=2 a_1=(4/ \pi) \int_{0}^\pi d Q \phi(Q,t) \sin(Q)$,
can be cast in the form:
\begin{equation}
i \dot c_a(t)=(\omega_a+ig)c_a(t)-\frac{4 \kappa_0}{\pi}
\int_{0}^{\pi}dQ \; \phi(Q,t) \sin(Q).
\end{equation}
By introducing the frequency $\omega$ of the continuum
\begin{equation}
\omega=-2 \kappa \cos(Q)
\end{equation}
and after setting
\begin{equation}
c(\omega,t)=-\sqrt{\frac{2}{\pi \kappa}} \phi(\omega,t)
\frac{1}{\left[1-\omega^2/(2 \kappa)^2 \right]^{1/4}},
\end{equation}
one finally obtains Eqs.(\ref{cme1a}) and (\ref{cme2a}) given in
the text.


\begin{thebibliography}{99}

\bibitem{Lang73}
R. Lang, O. Scully, and W.E. Lamb, Phys. Rev. A {\bf 7}, 1788
(1973).

\bibitem{Ching87}
S.C. Ching, H.M. Lai, and K. Young, J. Opt. Soc. Am. B {\bf 4},
1995 (1987).

\bibitem{Ching98}
E.S.C. Ching, P.T. Leung, A. Maassen van den Brink, W.M. Suen,
S.S. Tong, and K. Young, Rev. Mod. Phys. {\bf 70}, 1545 (1998).

\bibitem{Fano64}
U. Fano, Phys Rev.{\bf 124}, 1866 (1961).

\bibitem{Tannoudji}
C. Cohen-Tannoudji, J. Dupont-Roc, and G. Grynberg, {\it
Atom-Photon Interactions} (Wiley, New York, 1992).

\bibitem{Svelto}
O. Svelto, {\it Principles of Lasers}, fourth ed. (Springer,
Berlin, 1998).

\bibitem{note0}
It is remarkable as well that the usual gain/loss balance
condition for lasing threshold, with an exponential growth at the
onset of lasing, is valid even for less conventional laser
systems, such as in random lasers [see, for instance: V. S.
Letokhov, Sov. Phys. JETP {\bf 26}, 835 (1968); T. Sh.
Misirpashaev and C.W.J. Beenakker, Phys. Rev. A {\bf 57}, 2041
(1998); X. Jiang and C.M. Soukoulis, Phys. Rev. B {\bf 59}, 6159
(1999); A.L. Burin, M.A. Ratner, H. Cao, and S.H. Chang, Phys.
Rev. Lett. {\bf 88}, 093904 (2002)].

\bibitem{Piraux90}
B. Piraux, R. Bhatt, and P.L. Knight, Phys. Rev. A {\bf 41}, 6296
(1990).

\bibitem{Lai88}
H.M. Lai, P.T. Leung, and K. Young, Phys. Rev. A {\bf 37}, 1597
(1988).

\bibitem{Lewenstein88}
M. Lewenstein, J. Zakrzewski, T.W. Mossberg, and J. Mostowski, J.
Phys. B: At. Mol. Opt. Phys. {\bf 21}, L9 (1988).

\bibitem{John90}
S. John and J. Wang, Phys. Rev. Lett. {\bf 64}, 2418 (1990).

\bibitem{John94}
S. John and T. Quang, Phys. Rev. A {\bf 50}, 1764 (1994).

\bibitem{Kofman94}
A.G. Kofman, G. Kurizki, and B. Sherman, J. Mod. Opt. {\bf 41},
353 (1994).

\bibitem{Vats98}
N. Vats and S. John, Phys. Rev. A {\bf 58}, 4168 (1998).

\bibitem{Lambropoulos00}
P. Lambropoulos, G.M. Nikolopoulos, T.R. Nielsen, and S. Bay, Rep.
Prog. Phys. {\bf 63}, 455 (2000).

\bibitem{Wang03}
X.-H. Wang, B.-Y. Gu, R. Wang, and H.-Q. Xu, Phys. Rev. Lett. {\bf
91}, 113904 (2003).

\bibitem{Petrosky05}
T. Petrosky, C.-O. Ting, and S. Garmon, Phys. Rev. Lett. {\bf 94},
043601 (2005).

\bibitem{Tanaka06}
S. Tanaka, S. Garmon, and T. Petrosky, Phys. Rev. B {\bf 73},
115340 (2006).

\bibitem{Gaveau95}
B. Gaveau and L.S. Schulman, J. Phys. A: Math. Gen. {\bf 28}, 7359
(1995).

\bibitem{Villeneuve96}
P.R. Villeneuve, S. Fan, and J.D. Joannopoulos, Phys. Rev. B {\bf
54}, 7837 (1996).

\bibitem{Vahala03}
K.J. Vahala, Nature (London) {\bf 424}, 839 (2003).

\bibitem{Armani03}
D.K. Armani, T.J. Kippenberg, S.M. Spillane, and K.J. Vahala,
Nature (London) {\bf 421}, 925 (2003).

\bibitem{Asano04}
T. Asano and S. Noda, Nature (London) {\bf 429}, 6988 (2004).

\bibitem{Asano06}
T. Asano, W. Kunishi, B.-S. Song, and S. Noda, Appl. Phys. Lett.
{\bf 88}, 151102 (2006).

\bibitem{Painter99}
O. Painter, R. K. Lee, A. Yariv, A. Scherer, J. D. O'Brien, P. D.
Dapkus, and I. Kim, Science {\bf 284}, 1819 (1999).

\bibitem{Loncar02}
M. Loncar, T. Yoshie, A. Scherer, P. Gogna, and Y. Qiu, Appl.
Phys. Lett. {\bf 81}, 2680 (2002).

\bibitem{Park04}
H.G. Park, S.H. Kim, S.H. Kwon, Y.G. Ju, J.K. Yang, J.H. Baek,
S.B. Kim, and Y.H. Lee, Science {\bf 305}, 1444 (2004).

\bibitem{Altug05}
H. Altug and J. Vuckovic, Opt. Express {\bf 13}, 8819 (2005).

\bibitem{Fan98}
S. Fan, P.R. Villeneuve, J.D. Joannopoulos, and H.A. Haus, Phys.
Rev. Lett. {\bf 80}, 960 (1998); S. Fan, P.R. Villeneuve, J.D.
Joannopoulos, M.J. Khan, C. Manolatou, and H.A. Haus, Phys. Rev. B
{\bf 59}, 15882 (1999).

\bibitem{Xu00}
Y. Xu, Y. Li, R.K. Lee, and A. Yariv, Phys. Rev. E {\bf 62}, 7389
(2000).

\bibitem{Asano03}
 T. Asano, B.S. Song, Y. Tanaka, and S. Noda, Appl. Phys. Lett. {\bf 83}, 407 (2003).

\bibitem{Waks05}
E. Waks and J. Vuckovic, Opt. Express {\bf 13}, 5064 (2005).


\bibitem{Chak06}
P. Chak, S. Pereira, and J.E. Sipe, Phys. Rev. B {\bf 73}, 035105
(2006).


\bibitem{Fan05}
M.F. Yanik and S. Fan, Phys. Rev. A {\bf 71}, 013803 (2005).


\bibitem{LanLan05}
L.-L. Lin, Z.-Y. Li, and B. Lin, Phys. Rev. B {\bf 72}, 165330
(2005).


\bibitem{Stefanou98}
N. Stefanou and A. Modinos, Phys. Rev. B {\bf 57}, 12127 (1998).

\bibitem{Yariv99}
A. Yariv, Y. Xu, R.K. Lee, and A. Scherer, Opt. Lett. {\bf 24},
711 (1999).

\bibitem{Ozbay00}
M. Bayindir, B. Temelkuran, and E. Ozbay, Phys. Rev. Lett. {\bf
84}, 2140 (2000).

\bibitem{Olivier01}
S. Olivier, C. Smith, M. Rattier, H. Benisty, C. Weisbuch, T.
Krauss, R. Houdre, and U. Oesterle, Opt. Lett. {\bf 26}, 1019
(2001).

\bibitem{Liu05}
Y. Liu, Z. Wang, M. Han, S. Fan, and R. Dutton, Opt. Express {\bf
13}, 4539 (2005).

\bibitem{Nakazato96}
H. Nakazato, M. Namiki, and S. Pascazio, Int. J. Mod. Phys. B {\bf
10}, 247 (1996).

\bibitem{Regola}
P. Facchi and S. Pascazio, {\it La Regola d'Oro di Fermi}, in:
Quaderni di Fisica Teorica, edited by S. Boffi (Bibliopolis,
Napoli, 1999).

\bibitem{note1}
Note that, by extending the definition of $\Delta(\omega)$ outside
the interval $(\omega_1,\omega_2)$, the principal value of the
integral in Eq.(\ref{Omshift}) can be removed.

\bibitem{note2}
The value $\lambda_I$ ($\lambda_{II}$) defines the critical value
of coupling strenght above which a bound mode (discrete eigenvalue
of $H_0+H_{int}$) at frequency $\omega<\omega_1$
($\omega>\omega_2$) appears.

\bibitem{Yanik04}
M.F. Yanik and S. Fan, Phys. Rev. Lett. {\bf 92}, 083901 (2004).

\bibitem{Longhi06a}
S. Longhi, Phys. Rev. E {\bf 74}, 026602 (2006).

\bibitem{Longhi06b}
S. Longhi, Phys. Rev. Lett. {\bf 97}, 110402 (2006).

\bibitem{note3}
If gain saturation is accounted for and the dynamics may be
derived from a potential (e.g. after adiabatic elimination of
polarization and population inversion in the semiclassical laser
equations), the onset of laser oscillation is analogous to a
second-order phase transition [see, for instance: V. DeGiorgio and
M.O. Scully, Phys. Rev. A {\bf 2}, 1170 (1970); H. Haken, {\it
Synergetics}, second ed. (Springler-Verlag, Berlin, 1978)].



\bibitem{note4}
Denoting by $\mathcal{A}$ the matrix for the linear system (38) of
ordinary differential equations, the system is referred to as {\it
non-normal} whenever $\mathcal{A}$ does not commute with its
adjoint $\mathcal{A}^\dag$. One can show that transient energy
amplification is possible in an asymptotically-stable non-normal
system provided that the largest eigenvalue of
$\mathcal{A}+\mathcal{A}^\dag$ is positive (see e.g.
\cite{Farrell96}). Non-hermiticity is a necessary (but not
sufficient) condition to have transient energy grow in an
asymptotically-stable linear system.

\bibitem{Trefethen93}
L.N. Trefethen, A.E. Trefethen, S.C. Reddy, and T.A. Driscoll,
Science {\bf 261}, 578 (1993).
%
\bibitem{Farrell94}
B.F. Farrell and P.J. Ioannou, Phys. Rev. Lett. {\bf 72}, 1188
(1994).
%
\bibitem{Farrell96}
B. F. Farrell and P.J. Ioannou, J. Atmos. Sci. {\bf 53}, 2025
(1996).

\bibitem{Kartner99}
F.X. K\"{a}rtner, D.M. Zumb\"{u}hl, and N. Matuschek,  Phys. Rev.
Lett. {\bf 82}, 4428 (1999).


\bibitem{Longhi00}
S. Longhi and P. Laporta, Phys. Rev. E {\bf 61}, R989 (2000).


\bibitem{Firth05}
W.J. Firth and A.M. Yao, Phys. Rev. Lett. {\bf 95}, 073903 (2005).


\end{thebibliography}
\end{document}